\documentstyle[preprint,eqsecnum,aps]{revtex}

\begin{document}
\draft
\preprint{gr-qc/9401002}

\title{Another positivity proof and gravitational energy
       localizations}

\author{James M. Nester \thanks
       {Electronic address : nester@phyast.dnet.ncu.edu.tw}
   and  Roh-Suan Tung   \thanks
       {Electronic address : m792001@phy.ncu.edu.tw} }
\address{
     Department of Physics, National Central University \\
     Chung-Li, Taiwan 32054  }
\maketitle
\begin{abstract}
Two locally positive expressions for the gravitational
Hamiltonian, one using 4-spinors
the other special orthonormal frames, are reviewed. A
new quadratic 3-spinor-curvature identity is used
to obtain another positive expression for the Hamiltonian
and thereby
a localization of gravitational energy and positive energy
proof. These new
results provide a link between the other two methods.
Localization and prospects for quasi-localization are discussed.
\end{abstract}


\narrowtext
\section{Introduction}
For asymptotically flat gravitating systems
{\it total} energy is well defined and must be non-negative.
Each new positive total energy proof (e.g.\cite{Berg93})
offers some more insights.
Concerning the localization of the total energy,
although the equivalence principle
forbids a true local gravitational
energy density yet a suitable ``quasi-localization'' is desirable
\cite{Pen82}. A good candidate for a gravitational energy density
is the Hamiltonian density.  For asymptotically flat
Einstein gravity the Hamiltonian density has
the general form \cite{CG80}
\begin{eqnarray}
H(N) &=& \int d^3 x  \, 2 N^\mu
        G^0_{\,\mu} +\oint B        \nonumber\\
     &=& \int d^3 x \, \, N{\cal H} + N^k {\cal H}_k
              + \oint B \ ,  \label{eq:ADM-H}
\end{eqnarray}
which includes a boundary term at spatial infinity.  On a solution
the spatial integral vanishes, the {\it value} of the Hamiltonian,
$-16\pi\hbox{G} N^{\mu}p_{\mu}$, comes from the integral of the
boundary term over the 2-sphere at spatial infinity and determines
the total 4-energy momentum $p_\mu$.
The integrand of the boundary term, $B$, is only well defined up to
$O(r^{-2})$, moreover we have the freedom to choose the {\it lapse}
$N$ and {\it shift}  $N^k$. Together these allow a certain latitude
which can be exploited
to obtain a locally non-negative Hamiltonian density.
Indeed, such a form can be achieved in more than one way.

\section{The 4-covariant quadratic spinor Hamiltonian}

The first constructions of this type \cite{SCI} were  done
in the wake of the Witten positive energy proof.  It
was shown that the {\it Hamiltonian density}
for Einstein gravity could be
expressed as a 4-covariant quadratic spinor 3-form:
\begin{equation}
{\cal H}(\psi)
:=2\{D(\overline\psi\gamma_5\gamma)\wedge D\psi
-D\overline\psi\wedge D(\gamma_5\gamma\psi)\}\ .
 \label{eq:4Spinor-H}
\end{equation}
This remarkable result follows from: (i) the identity
\begin{eqnarray}
{\cal H}(\psi)\equiv&& 2 N^\mu G^\nu_{\,\mu} \eta_\nu
+d\{\overline\psi\gamma_5\gamma\wedge D\psi
+D(\overline\psi\gamma_5\gamma)\psi  \nonumber\\
   && -\overline\psi D(\gamma_5\gamma\psi)
   +D\overline\psi\wedge(\gamma_5\gamma\psi)\}
      \  ,     \label{eq:4SCI}
\end{eqnarray}
where
$N^\mu=\overline\psi\gamma^\mu\psi$
(for conventions see the appendix)
revealing that
${\cal H}(\psi)$ contains the appropriate projected
components of the Einstein
tensor (needed to generate the equations of motion) up to an exact
differential (which does not change the variational derivatives);
and (ii) the observation that $\delta\int{\cal H}(\psi)$ has an
asymptotically vanishing boundary integral since
${\cal H}(\psi)$ is asymptotically of order $O(r^{-4})$.

The Hamiltonian density (\ref{eq:4Spinor-H}) can be decomposed,
with respect to the normal to any
spacelike hypersurface, into positive and negative definite parts:
\begin{equation}
{\cal H}(\psi)  \simeq   4
    (g^{ab} D_a\psi^{\dag}  D_b\psi
       -  \mid \gamma^a D_a\psi \mid^2 )\,\eta_0  \  ,
\end{equation}
and is locally non-negative if
$\psi$ satisfies  the
Witten  equation (or certain modifications thereof)
\begin{equation}
   \gamma^a D_a \psi=0  \  ,
\end{equation}
 thereby permitting a non-negative
``localization'' of gravitational energy.

This mathematically elegant form of
the gravitational Hamiltonian has several virtues, in particular it
(i) is manifestly 4-covariant,
(ii) shows that total 4-momentum is future timelike,
(iii) can be evaluated on a spacelike surface extending to
future null infinity thereby showing
that the Bondi 4-momentum also is future timelike.
However it also has some liabilities, in particular
(a) the spinor field is physically mysterious,
(b) there is no direct relation to the customary variables,
(c) it yields an  unintuitive energy localization. (For the
Schwarzschild solution in isotropic cartesian frames
$\psi=(1+m/2r)^{-2}\psi_{\hbox{flat}}$
solves the Witten equation; using this result in the boundary
integral yields
1/8 of the total mass-energy inside the horizon and 7/8 outside.)
Consequently other Hamiltonian based positivity
proofs/localizations  were sought and found.

\section{The special orthonormal frame approach}
Another approach \cite{SOFPEP}  used orthonormal
frames and exploited their rotational gauge freedom.  The
ADM Hamiltonian (\ref{eq:ADM-H}) in an asymptotically
cartesian frame  has the form:
\begin{eqnarray}
      H(N) =&&   \int d^3 x \, \, N (
g^{-{1\over2}}(\pi^{mn}\pi_{mn}-{\textstyle{1\over2}}\pi^2)
- g^{1\over2} R  )     \nonumber\\
&&+2\pi^m{}_k\nabla_m N^k
 + \oint dS_k \,\,N\, \delta^{kc}_{ab}\,\Gamma^{ab}{}_c \ .
 \label{eq:ADM-B}
\end{eqnarray}
We choose $N^k=0$, use the divergence theorem to eliminate the
boundary term, parameterize the metric with orthonormal frames,
split the connection coefficients algebraically into a symmetric
tensor $q_{ab}$, a vector $q_c:=-\Gamma^a{}_{ca}$ and a scalar
$q:=\epsilon^{abc}\Gamma_{abc}$, and use the
{\it special orthonormal frame}
(SOF) \cite{SOF} rotational gauge conditions:
\begin{equation}
 q_k=4\partial_k\ln\Phi, \qquad q=\hbox{constant} ,
 \label{eq:SOF-gauge}
\end{equation}
to obtain the Einstein
Hamiltonian (i.e., energy) density in the form
\begin{eqnarray}
{\cal H}(N)&=& 8g^{1\over2}g^{nm}
\partial_n (N\Phi^{-1})\partial_m\Phi   \nonumber\\
&&+N\{ g^{-{1\over2}} (\pi^{ab}\pi_{ab}-{\textstyle{1\over2}}\pi^2)
+g^{1\over2} (q^{ab}q_{ab}-{\textstyle{1\over2}}q^2 ) \}  \  .
\label{eq:SOF-H}
\end{eqnarray}
This expression is good for both compact spatial surfaces (in which
case q is a non-vanishing constant) and for asymptotically
flat spatial  surfaces (in which case q vanishes).
For the latter case total energy is well defined;
a suitable choice of the lapse gives a positive total energy proof.

Many choices for the lapse  give a positive local energy density,
in  particular $N=\Phi^a, (a\ge1)$.
An especially attractive choice
is $N=\Phi$ which leads to the {\it gravitational energy density}
\begin{equation}
{\cal H}(\Phi)=\Phi
\{ g^{-{1\over2}}(\pi^{ab}\pi_{ab}-{\textstyle{1\over2}}\pi^2)
+g^{1\over2}(q^{ab}q_{ab}-{\textstyle{1\over2}}q^2 )\}  \ ,
\end{equation}
and the value
\begin{equation}
E=(16\pi\hbox{G})^{-1}\int_V {\cal H}(\Phi)\,d^3x
  = (2\pi\hbox{G})^{-1}\oint_S g^{1\over2}\nabla^k\Phi
dS_k    \ ,
\end{equation}
for the amount of energy localized
 within a volume V bounded by a surface S.
The value is non-negative for asymptotically flat maximal spatial
hypersurfaces. The {\it gravitational potential } $\Phi$
(generalized Newtonian
potential) satisfies the generalized Poisson equation:
\begin{equation}
   8g^{1\over2}\Delta\Phi={\cal H}(\Phi)
+16\pi \Phi g^{1\over2}\hbox{G}\rho \  .
\end{equation}
(This is just the
Hamiltonian
constraint and is essentially the scale equation of the usual
conformal approach to the Einstein initial value constraints
see, e.g., Choquet-Bruhat and York \cite{IV80} ).

This SOF Hamiltonian has certain virtues, especially
(i) the gauge conditions are conformally invariant so SOFs are
closely related to the usual variables of the standard
initial value constraints,
(ii) the oscillating physical modes are apparent in the SOF
gravitational energy density, and
(iii) the energy localization is physically reasonable; in
particular, all of the mass of the Schwarzschild solution
(note: $\Phi=(1+m/2r)^{-1}$ for the isotropic cartesian frame) is
within the horizon.
(Moreover there is some freedom here; the choice $N=\Phi^4$ produces
the same 1/8 inside the horizon as the 4-covariant spinor
Hamiltonian  for the Schwarzschild solution.)
However the SOF approach also has certain liabilities,
in particular
(a) the expression concerns only energy, it gives no restraint
on the momentum,
(b) the energy is guaranteed to be {\it locally} non-negative only
for maximal spacelike hypersurfaces,
(c) a maximal spacelike hypersurface cannot be extended to future
null infinity so this approach cannot give the Bondi mass-energy.

\section{A new 3-spinor proof and localization}
A link between the special orthonormal frame approach and the
4-covariant quad\-ratic spinor form of the Hamiltonian
has now been found in terms of a
new Hamiltonian based gravitational energy positivity proof
and localization which uses 3 dimensional spinors.

The key is a new spinor identity (see appendix)
\begin{eqnarray}
 2[\nabla( \varphi^{\dag}i\sigma) \wedge\nabla\varphi
 &-&\nabla \varphi^{\dag}\wedge\nabla(i\sigma\varphi)]\nonumber\\
 &&\equiv d B-(\varphi^{\dag}\varphi)\Omega^{ab}\wedge\zeta_{ab} \ ,
\label{eq:3SCI}
\end{eqnarray}
where
\begin{eqnarray}
B:=&&  \varphi^{\dag}i\sigma \wedge\nabla\varphi
-\varphi^{\dag}\nabla(i\sigma\varphi)  \nonumber\\
   &&   +\nabla(\varphi^{\dag}i\sigma)\varphi
   +(\nabla\varphi^{\dag})\wedge i\sigma\varphi \  .
\end{eqnarray}
Using this identity we replace the scalar curvature term,
$NRg^{1/2}d^3x=N\Omega^{ab}\wedge\zeta_{ab}$,
and the boundary term in the ADM Hamiltonian
(\ref{eq:ADM-B}) with the left hand
side of (\ref{eq:3SCI}) .
The Einstein Hamiltonian  (with $N=\varphi^{\dag}\varphi$,
$N^k=0$) can then be written as
\begin{eqnarray}
H(N) =&& \int \! d^3 x\,  (\varphi^{\dag}\varphi)
 g^{-{1\over2}}(\pi^{mn}\pi_{mn}
 -{\textstyle{1\over2}}\pi^2) \nonumber\\
 &&+ 2[\nabla( \varphi^{\dag}i\sigma) \wedge\varphi
 -\nabla \varphi^{\dag}\wedge\nabla(i\sigma\varphi)]
 \ . \label{eq:3Spinor-H}
\end{eqnarray}

An important property of the 3-spinor Hamiltonian
(\ref{eq:3Spinor-H}) is that
no additional boundary term at infinity is needed.
As Regge and Teitelboim \cite{RT74}
have nicely explained it is necessary that the boundary terms
in the variation of the Hamiltonian vanish asymptotically.
To verify this
property for (\ref{eq:3Spinor-H}) we need only check the
variation of the quadratic spinor terms (see appendix):
\begin{eqnarray}
\lefteqn{
\delta[\nabla(\varphi^{\dag}i\sigma)\wedge\nabla\varphi
-\nabla \varphi^{\dag}\wedge\nabla(i\sigma\varphi)] }\nonumber\\
&\equiv&-\Omega^{ab}\wedge\delta[(\varphi^{\dag}\varphi)\zeta_{ab}]
-{\textstyle{1\over2}} \delta\omega^{ab}\wedge\nabla
 [(\varphi^{\dag}\varphi)\zeta_{ab}] \nonumber\\
&&+ d \{ \delta(\varphi^{\dag}i\sigma)\wedge\nabla\varphi
 -\delta\varphi^{\dag}\nabla(i\sigma\varphi) \nonumber\\
&&\qquad +\nabla(\varphi^{\dag}i\sigma)\delta\varphi
 +(\nabla\varphi^{\dag})\wedge\delta(i\sigma\varphi) \} .
\end{eqnarray}
Since $\varphi\sim \hbox{constant} + O(1/r) $ and
$\delta\varphi \sim O(1/r)$
the boundary term falls off as $O(r^{-3})$, therefore vanishing
at spatial infinity for aymptotically flat initial data.

On a maximal hypersurface $ \pi=0 $, the kinetic terms in the
Hamiltonian (\ref{eq:3Spinor-H}) are non-negative.
The quadratic spinor terms can also be made non-negative.
Since the torsion vanishes the spinor terms reduce to
\begin{eqnarray}
\lefteqn{
2[ \nabla( \varphi^{\dag}i\sigma) \wedge\nabla\varphi
-\nabla \varphi^{\dag}\wedge\nabla(i\sigma\varphi)]}\nonumber\\
&=&  -4[ \nabla_a\varphi^{\dag} i\zeta^{abc}\sigma_c
  \nabla_b\varphi ]\zeta    \nonumber\\
&=&   4 [g^{ab} \nabla_a\varphi^{\dag} \nabla_b\varphi
 -\nabla_a\varphi^{\dag}\sigma^a\sigma^b \nabla_b\varphi]\zeta .
\end{eqnarray}
Hence
 the spinor terms are non-negative for
any asymptotically constant $\varphi$ satisfying
the 3-dimensional Dirac equation
\begin{equation}
   \sigma^a \nabla_a \varphi=0  \ . \label{eq:3Dirac}
\end{equation}
This is a linear elliptic equation similar to the Witten equation;
essentially the same arguments show that unique solutions exist.
 Hence the local energy density
is non-negative on asymptotically spacelike maximal slices.

This 3-spinor Hamiltonian approach by itself has
most of the liabilities of the other two approaches:
(a) it yields the same sort of unintuitive energy localization
as the 4-covariant spinor expression
(indeed they have identical values for  the
Schwarzschild solution but differ
when $K_{ab}\ne0\ne\overline\psi\gamma^a\psi$),
(b) the expression concerns only energy,
it gives no restraint on the momentum;
(c) the energy is guaranteed to be {\it locally} non-negative
only for {\it maximal} spacelike hypersurfaces,
(d) the maximal spacelike hypersurface cannot be extended to
future null infinity so it cannot give the Bondi mass-energy.
However, for the 3-spinor Hamiltonian some of the other 4-spinor
liabilities
are not so severe since  in this case (i) the spinor field
is not so mysterious, for (ii) there is a relation to the customary
variables via the SOF variables as we shall show below.  Indeed the
principal virtue of this approach is that it relates the other two
methods we have discussed.

\section{Relations between the methods}

The 3-spinor Hamiltonian expression (\ref{eq:3Spinor-H})
is intermediate between the 4-covariant spinor
Hamiltonian (\ref{eq:4Spinor-H}) discussed above and
the SOF Hamiltonian (\ref{eq:SOF-H}) .

On the one hand
it can be extracted as a piece of the 3+1
decomposition of the 4-covariant spinor Hamiltonian.
The orthonormal frame components of the metric compatible
4-connection  project into the components of the 3-connection
and the extrinsic curvature $K^a{}_b=-\Gamma^{0 a}{}_b$ hence
\begin{eqnarray}
  D_c\psi &=& \partial_c\psi
 - {\textstyle{1\over4}} \Gamma^{\alpha\beta}{}_c
 \gamma_{[\alpha}\gamma_{\beta]}\psi \nonumber\\
 &=&\partial_c\psi-{\textstyle{1\over4}}
 \Gamma^{ab}{}_c \gamma_{[a}\gamma_{b]}\psi
     -{\textstyle{1\over2}}\Gamma^{0b}{}_c
     \gamma_{[0}\gamma_{b]}\psi \nonumber\\
 &=&\nabla_c\psi
 +{\textstyle{1\over2}} K^b{}_c \gamma_{[0}\gamma_{b]}\psi .
\end{eqnarray}
Consequently the quadratic in $D\psi$ Hamiltonian decomposes into:
 (a)
quadratic terms in $K^a{}_b$ along with quadratic terms
in $\nabla\psi$ which are essentially
the 3-spinor Hamiltonian density (\ref{eq:3Spinor-H}),
 (b) linear terms
 in $K^a{}_b$ (they are of the form $2\pi^m{}_c\nabla_mN^c$ where
$N^c=\overline\psi\gamma^c\psi$)
which represent the momentum constraint.
(Note that the  3-spinor method decouples the spinor field
from $N^k$; this has both advantages and disadvantages.)

On the other hand the 3-spinor Hamiltonian not only
resembles the SOF approach in (i) using a vanishing shift, (ii)
considering the kinetic terms separately, (iii) relying on maximal
slices, and (iv) replacing the potential terms by an expression
using different variables, but,
moreover, there is a close relation
between the SOF variables and
spinor fields via  solutions to the 3 dimensional Dirac equation
(\ref{eq:3Dirac}).

Indeed the 3-dimensional Dirac equation explicitly
depends only on
the parts of the connection which appear in the gauge conditions
(\ref{eq:SOF-gauge})  :
\begin{eqnarray}
\sigma^c\nabla_c\varphi&=&\sigma^c(\varphi_{,c}+
{\textstyle{1\over 4}}
\Gamma^{ab}{}_c\sigma_{[a}\sigma_{b]}\varphi)\nonumber\\
&=&\sigma^c\varphi_{,c}-{\textstyle{1\over2}} q_b\sigma^b\varphi
+{\textstyle{1\over4}}
 i q\varphi  \  .
\end{eqnarray}
An asymptotically constant solution
to $\sigma^a\nabla_a\varphi=0$ can be factored into
a magnitude and a unitary transformation which determines an SOF
\cite{DM89} . Conversely, expressed
in terms of an SOF the Dirac equation reduces to
$\sigma^a\partial_a\Phi^{-2}\varphi=0$,
hence $\varphi=\Phi^2\varphi_{\hbox{const}}$.

\section{Localization and quasi-localization}

Our considerations have been concerned with
obtaining a positive {\it localization} of the
{\it total energy} by finding a good expression for the Hamiltonian
density.
Each localization depends on the solution to an elliptic
equation, which, in turn, depends on
the values on the boundary of the region.  Since the boundary
is at spatial infinity we can simply choose
suitable constant values as the physically appropriate
boundary conditions.

Beyond distributing the total
gravitational
energy, there is considerable
interest in ``quasi-localization'', i.e.,  determining the amount of
energy in a finite region {\it without} reference to what is outside.
The expressions we have discussed could also be used
for a finite region.  Then each of the (locally positive)
Hamiltonian densities provides a quasi-local energy.
The value of the positive quasi-local energy can be
obtained from the associated boundary integral.
The quasi-localization will
depend on the choice of {\it boundary
values} on the finite 2-surface bounding the region.
We, however, do not know how to decide which values
on a finite boundary
are a physically good choice.  Several
``quasi-localizations''
based upon 4-covariant
spinor expressions like (\ref{eq:4Spinor-H}) and
(\ref{eq:4SCI}) have been investigated by others\cite{DMB}.
Their methods of choosing boundary values for the 4-spinor field
could also be adapted
to our orthonormal frame or 3-spinor fields.
Canonical investigations
associated with a {\it finite}
region, with particular attention to the possible boundary terms and
their relation to
what is held fixed on the boundary, have been done for the
standard variables \cite{JKBY}.  Such a study
of the spinor or SOF parameterized Hamiltonian
should provide some guidance for the choice of appropriate
boundary values for finite regions.

Nontrivial examples of the
localizations  produced by the 4-spinor, SOF and 3-spinor techniques,
e.g., for the Kerr solution, would be instructive.
However, as noted, the localizations depend
on solving an elliptic system of equations, essentially the Dirac
equation.  Unfortunately, aside from the aforementioned spherically
symmetric case, there are hardly any
known exact solutions for the Dirac equation in curved spacetime
\cite{Shishkin}.

Forgoing direct comparisons for actual solutions we can compare the
expressions by considering desirable properties.
We know of no
gravitational energy localization method which is satisfactory.
One list \cite{Yao} for example, requires
(i) zero for flat spacetime, (ii) the standard value
for spherical solutions, (iii) the ADM value for an
asymptotically flat slice,
(iv) the Bondi value for an asymptotically null slice, (v) the
irreducible mass for the apparent horizon, (vi) positive and
monotonic.  Of the methods considered here,
the quadratic 4-spinor  expression certainly fails (ii) \& (v).  The
SOF Hamiltonian satisfies the positivity requirement (vi) only on
maximal slices, while
the maximal slice restriction precludes satisfying (iv).
The new 3 spinor technique has all of these failings.

\section{Conclusion}

We have presented a new positive total energy proof for
asymptotically flat Einstein gravity.  The proof uses a
3 dimensional spinor parameterization of the Hamiltonian and a new
3-spinor-curvature identity.  What insights has it yielded so far?

As a proof, considered on its own, this method has no
advantages and indeed
is less general than some other known proofs. More interesting is the
fact that it provides {\it another} independent method for
obtaining a {\it positive localization} of gravitational energy; yet
again, as a localization method, it has no apparent advantages.

Probably the most interesting thing is that it provides a {\it
link} between two other Hamiltonian based proofs and their associated
localizations.
At the very least
this link connects the somewhat mysterious
Witten spinor field proof and localization to
the more usual type variables.

Perhaps this link will play an essential role in finding
a modification of our expressions into a better
Hamiltonian density---one which permits a
positive energy proof and gravitational energy localization
combining the virtues of the  4-covariant spinor Hamiltonian and
special orthonormal frame approaches without the liabilities.
Such an expression, complimented by a good choice of boundary values
for finite regions would provide a physically reasonable
quasi-localization of gravitational energy.

\acknowledgments
This work was
supported by the National Science Council of the R.O.C. under
Contract No. NSC 82-0208-M-008-013.

\appendix
\section*{ Conventions and identities}

Our 4-dimensional conventions are:
metric signature $(-1,+1,+1,+1)$, orthonormal coframe $\theta^\mu$,
unit volume element $\eta$, unit 3-form basis $\eta_\mu=*\theta_\mu$.
Spinor conventions
$\gamma_5=\gamma^0\gamma^1\gamma^2\gamma^3$,
$$\gamma_\mu\gamma_\nu+\gamma_\nu\gamma_\mu
=-2g_{\mu\nu}=2\hbox{diag}(+1,-1,-1,-1),$$
$\gamma=\gamma_\mu\theta^\mu$ and
$D\psi=d\psi-{1\over4}
\omega^{\mu\nu}\gamma_{[\mu}\gamma_{\nu]}\psi$ is the
covariant differential.

The 3-dimensional spinor conventions used are:
$\sigma=\sigma_c\theta^c$ where
$$\sigma_a\sigma_b+\sigma_b\sigma_a=2\delta_{ab},$$
with $\sigma_{ab}={1\over 4} [\sigma_a,\sigma_b], $ so
$\sigma_c\sigma_{ab}+\sigma_{ab}\sigma_c=i\zeta_{abc},$ and
$\zeta_{ab}=\zeta_{abc}\theta^c,$
where $\zeta_{abc}$ is the  3-dimensional Levi-Civita tensor with
$\zeta_{123}=+1$.

The identity connecting  the 3-dimensional scalar curvature to the
spinor expression in the Hamiltonian can be verified as follows:
\widetext
\begin{eqnarray}
\lefteqn{
 2[\nabla (\varphi^{\dag}i\sigma)\wedge\nabla\varphi
 -\nabla \varphi^{\dag}\wedge\nabla(i\sigma\varphi)]}\nonumber\\
 &\equiv& d B
 -[ -\varphi^{\dag}i\sigma\nabla^2\varphi
 +\nabla^2(\varphi^{\dag}i\sigma)\varphi
    -\varphi^{\dag}\nabla^2(i\sigma\varphi)
    +(\nabla^2\varphi^{\dag})i\sigma\varphi  ] \nonumber\\
 &\equiv& d B -{\textstyle{1\over 2}} \Omega^{ab}\wedge [
    -\varphi^{\dag}i\sigma\sigma_{ab}\varphi
    -\varphi^{\dag}i\sigma\sigma_{ab}\varphi
    -\varphi^{\dag}i\sigma_{ab}\sigma\varphi
    -\varphi^{\dag}i\sigma_{ab}\sigma\varphi  ] \nonumber\\
 &\equiv& d B +  \Omega^{ab} \wedge
   [\varphi^{\dag}(i\sigma\sigma_{ab}+i\sigma_{ab}\sigma)\varphi]
\equiv d B
    -(\varphi^{\dag}\varphi) \Omega^{ab} \wedge \zeta_{ab},
\end{eqnarray}
where $B:= \varphi^{\dag}i\sigma\wedge\nabla\varphi
-\varphi^{\dag}\nabla(i\sigma\varphi)
        +\nabla(\varphi^{\dag}i\sigma)\varphi
        +(\nabla\varphi^{\dag})\wedge i\sigma\varphi $.

Similarly, we calculate the variation of the quadratic spinor terms
\begin{eqnarray}
\lefteqn{
\delta[\nabla (\varphi^{\dag}i\sigma)\wedge\nabla\varphi
-\nabla \varphi^{\dag}\wedge\nabla(i\sigma\varphi)]} \cr
&\equiv&[ \nabla\delta(\varphi^{\dag}i\sigma)\wedge\nabla\varphi
-\nabla\delta\varphi^{\dag}\wedge\nabla(i\sigma\varphi)
 +\nabla(\varphi^{\dag}i\sigma)\wedge\nabla\delta\varphi
-(\nabla\varphi^{\dag})\wedge\nabla\delta(i\sigma\varphi)] \cr
&&+[ (\delta\nabla)(\varphi^{\dag}i\sigma)\wedge\nabla\varphi
-(\delta\nabla)\varphi^{\dag}\wedge\nabla(i\sigma\varphi)
+\nabla(\varphi^{\dag}i\sigma)\wedge(\delta\nabla)\varphi
-\nabla\varphi^{\dag}\wedge(\delta\nabla)(i\sigma\varphi)]\cr
&\equiv& d [\delta(\varphi^{\dag}i\sigma)\wedge\nabla\varphi
-\delta\varphi^{\dag}\nabla(i\sigma\varphi)
    +\nabla(\varphi^{\dag}i\sigma)\delta\varphi
    +(\nabla\varphi^{\dag})\wedge\delta(i\sigma\varphi) ]\cr
&& - [-\delta(\varphi^{\dag}i\sigma)\wedge\nabla^2\varphi
-\delta\varphi^{\dag}\nabla^2(i\sigma\varphi)
    +\nabla^2(\varphi^{\dag}i\sigma)\delta\varphi
    +\nabla^2\varphi^{\dag}\wedge\delta(i\sigma\varphi)]\cr
&&+{\textstyle{1\over2}}\delta\omega^{ab}\wedge
[- (\varphi^{\dag}i\sigma)\sigma_{ab}\wedge\nabla\varphi
+\varphi^{\dag}\sigma_{ab}\nabla(i\sigma\varphi)
 + \nabla(\varphi^{\dag}i\sigma)\sigma_{ab}\varphi
 +\nabla\varphi^{\dag}\sigma_{ab}\wedge i\sigma\varphi ]\cr
&\equiv&  d [\dots] -{\textstyle{1\over2}}\Omega^{ab}
\wedge[-\delta(\varphi^{\dag}i\sigma)\sigma_{ab}\varphi
-\delta\varphi^{\dag}\sigma_{ab}(i\sigma\varphi)
   -(\varphi^{\dag}i\sigma)\sigma_{ab}\delta\varphi
   -\varphi^{\dag}\sigma_{ab}\delta(i\sigma\varphi) ]\cr
&&+{\textstyle{1\over2}}\delta\omega^{ab}\wedge
\nabla[ \varphi^{\dag}(i\sigma\sigma_{ab}
+\sigma_{ab}i\sigma)\varphi]\cr
&\equiv&  d [\dots]
+ \Omega^{ab}\wedge\delta
[ \varphi^{\dag}(i\sigma\sigma_{ab}+\sigma_{ab}i\sigma)\varphi]
 +{\textstyle{1\over2}}  \delta\omega^{ab} \wedge\nabla
 [ \varphi^{\dag}(i\sigma\sigma_{ab}
     +\sigma_{ab}i\sigma)\varphi]    \cr
&\equiv&  d [\dots] - \Omega^{ab} \wedge\delta
[ (\varphi^{\dag}\varphi)\zeta_{ab} ]
-{\textstyle{1\over2}} \delta\omega^{ab} \wedge\nabla
[ (\varphi^{\dag}\varphi)\zeta_{ab} ].
\end{eqnarray}

We have recently discovered \cite{NTZ}
that there are many identities like (\ref{eq:4SCI}) and
(\ref{eq:3SCI}) in Riemann or Riemann-Cartan spaces
of any dimension.
\narrowtext

\end{document}